\documentstyle[]{l-school}
\begin{document}
\markboth{S.P. Klevansky et al.}{DEVELOPING TRANSPORT THEORY ...}
\setcounter{part}{10}
%
\title{Developing Transport Theory to Study the Chiral Phase Transition}
\author{S.P. Klevansky, A. Ogura, P. Rehberg and J. H\"ufner}
\institute{Institut f\"ur Theoretische Physik, Philosophenweg 19, \\
D-69120 Heidelberg, Germany
}
\maketitle
\section{INTRODUCTION}
One of the fundamental questions that is open to both theory and experiment
is how to observe the chiral phase transition.   Indications that a phase
transition from a chirally symmetric to a chirally broken phase have been
obtained a long time ago:   lattice gauge simulations of quantum
chromodynamics (QCD), which are equilibrium calculations that can include
only temperature but not finite density, show a phase transition at a 
finite critical temperature $T_c$ \cite{qm}.    Unfortunately, due to the 
difficulties inherent in a confining theory such as QCD, it is impossible
to observe this transition directly experimentally, let alone obtain
experimental values for the usual quantities that are associated with 
critical phenomena, such as critical exponents.

The experimental search for clues to a phase transition that has been 
conceived, is to proceed via heavy ion collisions, in which both high 
temperatures and densities are to be reached.   Much effort has been 
expended in this direction and will continue to be so during the early
course of the next century, particularly at RHIC and CERN.    However,
from the theoretical side, there is at present no single conclusive
observable that might indicate the occurrence of the chiral phase 
transition.   In addition, the simple equilibrium picture of the static 
lattice simulations is inadequate for describing heavy ion collisions.
Thus we may speculate as to whether a concrete signal of this phase 
transition can be obtained from a non-equilibrium formulation of the problem
or transport theory, that is based on an underlying chiral Lagrangian.
Here one may surmise whether phenomena such as critical scattering, i.e.
divergence of the scattering cross-sections at the phase transition, may
leave an obvious signal, or simply the fact that the dynamically generated
quark mass is a function of space and time may leave some visible remnant
of a phase transition.

Thus our goal is to construct a consistent transport theory that is based on
the Nambu--Jona-Lasinio  (NJL) model, which to order $1/N_c$ also includes
the mesonic degrees of freedom and the hadronization of quarks into these
degrees of freedom.   In this report, we discuss primarily the first term
in the $1/N_c$ expansion in the collision, explicitly relating this to 
Feynman diagrams for quark-quark and quark-antiquark scattering (for
details, see Ref.\cite{ogu}), leaving the mesonic sector to a future publication 
\cite{peter}. 
\section{GENERAL TRANSPORT THEORY FOR FERMIONS.}

Non-equilibrium phenomena are completely described by the 
Schwinger - Kel\-dysh formalism for Green functions \cite{schw}.
   Many conventions
exist in the literature.   For our purposes, it is simplest to use
the convention of Landau \cite{lif}.   In this, the designations $+$ and
$-$
attributed to the closed time path that is shown in Fig.1, and the
fermionic Green functions are defined as
\begin{eqnarray}
iS^c(x,y) &=& \left <T\psi(x) \bar \psi(y)\right > = iS^{--}(x,y)\nonumber
\\
iS^a(x,y) &=& \left <\tilde T\psi(x)\bar\psi(y)\right > = iS^{++}(x,y)
\nonumber \\
iS^{>}(x,y) &=& \left <\psi(x)\bar\psi(y)\right > = iS^{+-}(x,y) \nonumber
\\
iS^<(x,y) &=& -\left <\bar\psi(y)\psi(x)\right > = iS^{-+}(x,y).
\end{eqnarray}
In a standard fashion, one
constructs equations of motion for the matrix of Green functions and one
moves to relative and centre of mass variables, $u=x-y$ and $X=(x+y)/2$.  A
Fourier transform with respect to the relative coordinate, or Wigner
transform, \begin{equation} S(X,p) = \int d^4u e^{ip\cdot u} S\left(X+\frac
u2,X-\frac u2\right) \end{equation} is then performed.  Of particular
interest is
the equation of motion for $S^{-+} =S^<$.  Regarding this, together with the
equation of motion for the adjoint function $S^{<\dagger}$, one arrives at
the so-called transport and constraint equations by adding and subtracting
these.  One finds 
\begin{equation}
 \frac{i\hbar}2\{\gamma^\mu,\frac{\partial
S^{-+}}{\partial X^\mu}\} + [\not p,S^{-+}(X,p)] =
I_{-}
\label{full1}
\end{equation} 
and
\begin{equation}
 \frac{i\hbar}2[\gamma^\mu,\frac{ \partial S^{-+}}{\partial
X^\mu}] + \{ \not p - m_0,S^{-+}(X,p)\} = I_{+},
\label{full2}
\end{equation}
respectively.  Here
\begin{equation}
I_{\mp} = I_{{\rm coll}} + I^A_{\mp} + I^R_{\mp},
\label{is}
\end{equation}
where the collision term is
\begin{eqnarray}
I_{{\rm coll}} &=&
\Sigma^{-+}(X,p) \hat\Lambda S^{+-}(X,p) - \Sigma^{+-}(X,p) \hat\Lambda
S^{-+}(X,p) \nonumber \\
&=& I_{{\rm coll}}^{{\rm gain}} - I_{{\rm coll}}^{{\rm loss}},
\label{collision}
\end{eqnarray}
and terms containing retarded (R) and advanced (A) components are contained
in
\begin{equation}
I^R_{\mp} =  - \Sigma^{-+} ( X,p ) \hat \Lambda
                                  S^{\rm R} ( X,p )
         \pm S^{\rm R} ( X,p ) \hat\Lambda
                                         \Sigma^{-+} ( X,p )
\label{irmp}
\end{equation}
and
\begin{equation}
I^A_{\mp} =\Sigma^{\rm A} ( X,p ) \hat\Lambda
                                  S^{-+} ( X,p )
         \mp S^{-+} ( X,p ) \hat\Lambda
                                      \Sigma^{\rm A} ( X,p ).
\label{iamp}
\end{equation}
In these equations, $\Lambda$ is given as
\begin{equation}
\hat\Lambda = \exp\left(-\frac{i\hbar}2(
\frac{\overleftarrow\partial}{\partial X^\mu}\frac
{\overrightarrow\partial}{\partial p_\mu} - \frac{
\overleftarrow\partial}{\partial p_\mu} \frac{\overrightarrow\partial}
{\partial X^\mu})\right).
\label{hatlambda}
\end{equation}
Note that the equations (\ref{full1}) and (\ref{full2}) are
general equations that are generic for {\it any}
relativistic fermionic theory
\cite{heinz}.

\section{APPROXIMATION SCHEMES - APPLICATION TO THE NJL MODEL}

A double expansion in inverse powers of the number of colors $1/N_c$ and
$\hbar$ is performed.     We explain this briefly in the context of the 
model.

The simplest form of the NJL Lagrangian 
\begin{equation}
 {\cal L} =  \overline{\psi} \left( i \hbar \not\!\partial -
                                                    m_0 \right) \psi
            + G \left[ \left( \overline{\psi} \psi \right)^{2}
                 + \left( \overline{\psi} i \gamma^{5} \psi \right)^{2}
                  \right]~,
\label{lagrange}
\end{equation}
is considered.   Here $G$ is a coupling strength and $m_0$ is the current
quark mass.   The color degree of freedom is implicit.   As this is a strong
coupling theory, an expansion in $G$ is inadmissable, and the usual
technique
involves an expansion in $1/N_c$.    In this ordering, the lowest order
Feynman graph is a Hartree term, which describes the mean field 
experienced by the quarks, while higher orders introduce collisions.   We
examine these two cases individually.   In order to make connection with the
classically known transport equations, we consider an expansion of the
transport and constraint equations in powers of $\hbar$.

  In the  transport and
constraint equations, Eqs.(\ref{full1}) and (\ref{full2}), the right
hand sides $I_-$ and $I_+$, respectively depend  on $\hbar$ in two ways:
(i) The derivative operator $\hat \Lambda$, Eq.(\ref{hatlambda}), contains
$\hbar$ to all orders.    Its expansion in $\hbar$ generates the various
terms of the so-called gradient expansion, and (ii) the self-energies
$\Sigma^R$ and $\Sigma^{-+}$ may have overall factors in $\hbar$.    While   
to order $O((1/N_c)^0)$, $\Sigma^R = m$, and is directly related to the   
mass, which we treat as a quantity that has a direct classical
interpretation, we find $\Sigma^{-+}\propto\hbar d\sigma/d\Omega$, and is
only
related via a factor $\hbar$ to a quantity with direct physical  
interpretation.   Note that while we use the overall factors in $\hbar^0$ or
$\hbar^1$ in $\Sigma^R$ or $\Sigma^{-+}$ in our $\hbar$ classification,
we do not expand the quantities $m$ or $d\sigma/ d\Omega$ in powers of
$\hbar$.    We will detail this strategy in the calculation of the mean
field term and collision integrals.                 
\subsection{The Hartree Approximation}
The leading term in the $1/N_c$ expansion corresponds to the Hartree
approximation.   The Hartree self-energy can be evaluated, given a form for
the Green function.   Here we assume that
\begin{eqnarray}
S^{-+}_H(X,p) &=& 2\pi i \frac 1{2E_p} [\delta(p_0-E_p)\sum_s u_s(p) \bar u_s(p)
f_q(X,p) \nonumber \\
 &&\qquad + \delta(p_0 + E_p) \sum_s v_s(-p) \bar v_s(-p) \bar f_{\bar q}(X,-p)
\label{quasi}
\end{eqnarray} 
with $E^2_p = p^2 + m^2(X)$ and $f_{q,\bar q}$ the (spin independent) quark
and antiquark distribution functions.     This is the quasiparticle
assumption.   Similar expressions can be written down for the other Green
functions of Eq.(1).
Inserting the quasiparticle {\it ans\"azte}
 into the Wigner transformed expression for the
self-energy in the Hartree approximation, $-i\Sigma_H^{--} = 2i\hbar {\rm
tr} iS_H^{--}(x,x)$ leads to the non-equilibrium gap equation
\begin{equation}
m(X) = m_0 + 4GN_cm(X)\int\frac{d^3p}{(2\pi\hbar)^3} \frac 1{E_p(X)} [ 1 -
f_q(X,p)
- f_{\bar q}(X,p)].
\label{gap}
\end{equation}
Some manipulations are now necessary for the evaluation of (3) and (4). 
The
evaluation of $I_{\pm}$ is made according to the $\hbar$ expansion.
Then, on performing a spinor trace and integrating over positive 
or negative energies, leads one to the Vlasov equation
\begin{equation}
p^\mu\partial_\mu f_{q,\bar q}(X,\vec p) + m(X)\partial_\mu
m(X)\partial_p^\mu
f_{q,\bar q}(X,\vec p) = 0,
\label{vlasov}
\end{equation}
where $p^0=E_p(X)$, and the constraint equation
\begin{equation}
(p^2 - m^2(X))f_{q,\bar q} = 0.
\label{hatconstr}
\end{equation}
to leading order in $\hbar$.   Note that the constraint equation validates
the quasiparticle {\it ansatz}, and that the equations Eq.(\ref{vlasov})
and (\ref{hatconstr}) can be solved self-consistently.

\subsection{The Collision Term}
There are several graphs which contribute to the next order in the $1/N_c$
expansion.  One of these is displayed in Fig.~1(a).   The accompanying
graph of Fig.~1(b) is of higher order in $1/N_c$, and is not required on
this basis, but, as it turns out, is essential for a complete and
correct identification of Feynman amplitudes in casting the collision 
term into a Boltzmann-like form.   The initial task is the evaluation of
$I_{{\rm coll}}$ of Eq.(6), with scalar or pseudoscalar vertices.   There
are three possible combinations, $\Sigma_\sigma$, $\Sigma_\pi$ and
$\Sigma_{{\rm coll}}$, which contain unity and $i\gamma_5$ throughout or
unity or $i\gamma_5$ in a mixed fashion.   For example, the scalar 
self-energy is determined to be
\begin{eqnarray}
\Sigma_\sigma^{+-}(X,p) &=& - 4G^2\hbar^2\int \frac{d^4p_1}{(2\pi\hbar)^4}  
\frac{d^4p_2}{
(2\pi\hbar)^4}\frac{d^4p_3}{(2\pi\hbar)^4}
(2\pi\hbar)^4\delta(p-p_1+p_2-p_3) \nonumber \\
      &\times& [ S^{+-}(X,p_1) {\rm tr}\left( S^{-+}(X,p_2) S^{+-}(X,p_3)
\right)
\nonumber \\
& &-S^{+-}(X,p_1)S^{-+}(X,p_2)S^{+-}(X,p_3)  ],
\label{sigsig}
\end{eqnarray} 
while the other self-energies are similar \cite{ogu}.

\begin{figure}                                           
\vspace{4cm}
\caption{Contributions to the self-energy that are of higher order
 in $1/N_c$ than
the Hartree term}                                        
\label{fig1}                                             
\end{figure}         

As before, performing a spinor trace and integrating over a positive or
negative energy range is required, so that the loss term, for example, is
correspondingly given as
\begin{equation}
J^{q,{\rm loss}}_{{\rm coll},\sigma} = \frac{2\pi i}{2E_p}{\rm tr}\left[
\Sigma^{+-}_\sigma(X,p_0=E_p,\vec p) \sum_s u_s(p)\bar u_s(p) f_q(X,\vec p)
\right],
\label{whole}
\end{equation}
which, when evaluated using  the quasiparticle approximations for the Green
functions, leads to eight terms.   After some algebra, one can show that
only terms that represent elastic scattering survive energy integration.
Other terms that contain vacuum fluctuations, as well as particle production
and annihilation are present at this level, see Fig.~2,  but are 
suppressed by the quasiparticle approximation.   An off-shell {\it ansatz}
is required for them not to vanish.     It is also important to note that
both diagrams of Fig.~1 are essential for the identification of the Feynman
amplitudes for elastic scattering in the $s$, $t$ and $u$ channels, as
appropriate.    While graph (a) leads to the square of the individual
amplitudes, graph (b) is necessary for the mixed terms that allow one to
construct a matrix element squared that involves several channels.
     The net result is
\begin{eqnarray}
&&p^\mu\partial_\mu f_q(X,\vec p) + m(X)\partial_\mu m(X)\partial_p^\mu
f_q(X,\vec p) = \nonumber \\
& & N_c\int d\Omega \int \frac {d^3p_2}{(2\pi\hbar)^3 2E_{p_2}} |\vec v_p-\vec
v_2|
2E_p 2E_{p_2}
\nonumber \\
&\times&\{ \frac 12
\frac{d\sigma}{d\Omega}|_{qq\rightarrow qq}(p2\rightarrow 13)(f_q(p_1)\bar
f_q(p_2) f_q
(p_3) \bar f_q(p) - \bar f(p_1) f_q(p_2) \bar f_q(p_3) f_q(p)) \nonumber \\
& & +\frac{d\sigma}{d\Omega}|_{q\bar q\rightarrow q\bar q}(p2\rightarrow 13)
(f_q(p_1) \bar             
f_{\bar q}(p_2) f_{\bar q}(p_3) \bar f_q(p)
- \bar                     
f_q(p_1) \bar f_{\bar q}(p_3) f_{\bar q}(p_2) f_q(p))\}, \nonumber \\
\label{boltmann}           
\end{eqnarray}           
where the scattering cross-sections contain $s$, $t$ and $u$ channel graphs
as appropriate, in the Born approximation and 
which is the relativistic generalization of the result of Kadanoff and Baym.
To lowest order in $\hbar$, arguments can be given to suggest that the
constraint equation remains unaltered.

\begin{figure}
\vspace{7cm}
\caption{Processes of the form $q\rightarrow (q\bar q)q$ and $(q\bar
q)
\rightarrow q$, as well as vacuum fluctuations that are suppressed in 
the quasiparticle approximation.   These are heuristic graphs and are not
Feynman graphs}
\label{fig2}
\end{figure}

\section{DISCUSSION}
The most important facets of this calculation are (i) in identifying precise
Feynman amplitudes that contribute to the scattering processes in the
collision term, and which represents the relativistic generalization of the
Kadanoff-Baym work \cite{kaba}, and (ii) the recognition of additional processes
that would be present should a quasiparticle assumption be relaxed, such
as vacuum fluctuations and particle production and annihilation.  The
double expansion in $\hbar$ and $1/N_c$ lays the foundation for including
additional higher order terms in the $1/N_c$ expansion, from which mesons
are constructed, and which enables one to build a dynamical model for both
mesons and quarks.   For this, we refer the interested reader to
\cite{peter,hirschegg}.

We have started numerical simulations of this chiral transport theory.  A
first study that solves the Vlasov equation by means of a finite difference
technique and which inserts  collision terms in the 
relaxation time approximation for three flavors is presented in Ref.
\cite{peter2}.  In this, one finds that the collisions determine the
high energy parts of the particle distribution, which is exponential.  
The mean field, on the other hand, produces an enhancement over the
exponential fits at low energies.   The behavior of the quark mass as
a function of time is such that a dip occurs in the initial state, and this
gradually flattens as the quarks flow out.
\ack{This work has been supported in part by the Deutsche
Forschungsgemeinschaft DFG under the contract number Hu 233/4-4, and by the
German Ministry for Education and Technology under contract number 
06 HD 742.}
%

%
%
%
%
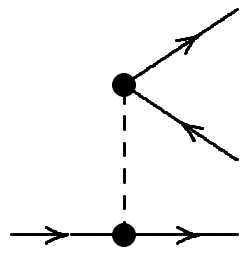
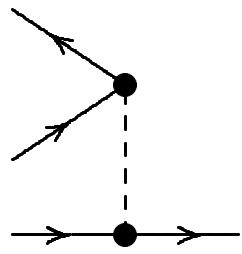
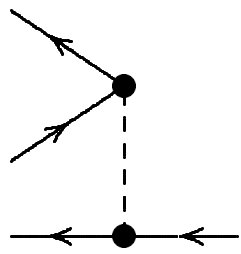
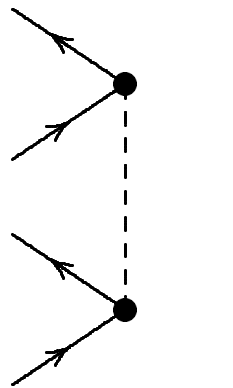
\end{document}